\documentclass[9pt,technote]{IEEEtran}

\usepackage[T1]{fontenc}		
\usepackage[english]{babel}	
\usepackage[english]{varioref}	

\usepackage{listings}		
\usepackage{xcolor}
\usepackage{url}			
\usepackage{microtype}		
\usepackage{import}			
\usepackage{amsmath}		
\usepackage{amssymb}
\usepackage{amsthm}
\usepackage{pgfplots}		
\usepackage{filecontents}	
\usepackage{braket}			
\usepackage{tikz}			
\usetikzlibrary{quantikz}
\usepackage{blochsphere}	

\usepackage[autostyle,italian=guillemets]{csquotes}  
\usepackage[backend=biber, sorting=none]{biblatex}

\usepackage[font=small]{caption}	
\usepackage{subfig}			
\usepackage{siunitx}		
\usepackage{sidecap}		

\usepackage{hyperref}		

\newcommand{\gateH}{\textbf{H}}

\newcommand{\gateI}{\textbf{I}}

\begin{document}

\title{A Quantum Edge Detection Algorithm}
\author{Giacomo Cavalieri*, Dario Maio*\\
*Department of Computer Science and Engineering, University of Bologna, Italy}
\maketitle

\begin{abstract}
The application of quantum computing to the field of image processing has produced several promising applications: quantum image representation techniques have been developed showing how, by taking advantage of quantum properties like entanglement and superposition, many image processing algorithms could have an exponential speed-up in comparison to their ``classical'' counterparts. 

In this paper, after briefly discussing some of the main quantum image representation methods, we propose an improved version of a quantum edge detection algorithm.
\end{abstract}

\begin{IEEEkeywords}
Quantum Edge Detection, Quantum Computing
\end{IEEEkeywords}

\section{Introduction}
Digital image processing, paired with the advances in artificial intelligence, has gained more and more relevance and has provided interesting applications: ranging from medicine \parencite{cit:skin-cancer} to self-driving cars. However, the increasing volume of data to process and the complexity of many image processing algorithms may pose a threat to future progress. Quantum computing could help overcome these limitations in a way classical computers could not with algorithms exponentially faster than their ``classical'' counterparts.

Over the years the discipline of quantum computing has proved many times how, by leveraging quantum properties like entanglement and superposition, it could solve problems with an exponential speed-up in comparison to classical solutions: two of the most prominent examples are the Shor Factorization Algorithm \parencite{cit:shor} and the Grover search algorithm \parencite{cit:grover}.
\\
\\
In Section \ref{sec:qimr} we briefly discuss some of the main Quantum Image Representation methods. 
In Section \ref{sec:qed}, after having analysed a quantum edge detection algorithm we propose an improved version. 
In Section \ref{sec:case} is shown a simple example of the application of this algorithm in the comparison of images. 
Lastly, in Section \ref{sec:concl} we report our conclusions, highlighting some of the challenges to current research and experimentation.
  

\section{Quantum Image Representation}\label{sec:qimr}
\subsection{FRQI}

\subsubsection{Image encoding}
FRQI (\textit{Flexible Representation for Quantum Images}) is a method to encode an image as a superposition of computational basis quantum states. It can be used on $2^{n}\times2^{n}$ images, requiring $2n+1$ qubits:
\begin{equation} \label{eq:FRQI-state}
	\ket{I} = \frac{1}{2^{n}}\sum_{i=0}^{2^{2n}-1}(\cos{\theta_{i}}\ket{0} + \sin{\theta_{i}}\ket{1})\otimes\ket{i},
\end{equation}
where $\theta_{i} \in [0,\pi/2]$ encodes the pixel intensity and $\ket{i}$ its position inside the image. The quantum state $\ket{I}$, as defined in \eqref{eq:FRQI-state}, is already normalized:
\begin{equation}
	||\ket{I}|| = \frac{1}{2^{n}}\sqrt{\sum_{i=0}^{2^{2n}-1}(\cos^{2}{\theta_{i}}+\sin^{2}{\theta_{i}})} = \frac{1}{2^{n}}\sqrt{2^{2n}}=1.
\end{equation}

One of the main advantages of this method is the low number of qubits required to encode the image: a square image with $2^{n}\times2^{n}$ can be encoded with just $2n+1$ qubits. Moreover, as the intensity data is encoded in a single qubit, pixel transformations (known as CTQI) can be implemented by applying a single quantum gate \parencite{cit:yan2014}.

\subsubsection{Image retrieval}
The measurement of a quantum circuit encoding an image with the FRQI method yields a state of the form $\ket{c}\ket{i}_{2n}$, where $\ket{c}$ is the single qubit used to encode the intensity data while $\ket{i}_{2n}$ encodes its position inside the image. However, a single measurement cannot give any useful information about the pixel intensity since it is encoded in the state amplitude. In order to get an approximation of the amplitude it is necessary to perform many measurements: given a generic state $\ket{c}\ket{i}_{2n}$ it is useful to define $N_{0}$ and $N_{1}$. They respectively indicate the times the states $\ket{0}\ket{i}_{2n}$ and $\ket{1}\ket{i}_{2n}$ are measured.
The probability of measuring the state $\ket{0}\ket{i}_{2n}$ can then be approximated as:
\begin{equation}
	P(\ket{0}\ket{i}_{2n}) = \cos^{2}(\theta/2) = \frac{N_{0}}{N_{0}+N_{1}}.
\end{equation}
This way an approximation of $\theta$ of the pixel in the position encoded by $\ket{i}_{2n}$ can be computed.
\\
\\
This approach presents a problem: to recover an accurate approximation of the intensity of each pixel, it would be necessary to perform a sufficiently high number of measurements for each of the $2^{2n+1}$ states. However, the exact image cannot be extracted from the circuit.
\subsection{NEQR}
The FRQI model leverages the quantum superposition of states to represent images. However, since the color data is encoded in the amplitude of each state it is impossible --as shown in the preceding paragraph-- to exactly recover the original image. NEQR (\textit{Novel Enhanced Quantum Representation}) is a way to encode images proposed in 2013 in \parencite{cit:zang}. It still keeps the base idea of FRQI of using quantum superposition of states, but provides an improvement in the image retrieval phase.

\subsubsection{Image encoding}
NEQR still uses quantum states to encode the position of pixels inside an image. The main difference with FRQI is that it encodes the pixel intensity in a further sequence of qubits entagled with the one representing the position. A grayscale image $I$ with size $2^{n}\times2^{n}$ and gray range from $0$ to $2^{q}-1$ --usually $2^{q}-1=255$-- can be encoded in the following quantum state of $2n+q$ qubits:
\begin{equation}\label{eq:NEQR-state}
	\ket{I} = \frac{1}{2^{n}}\sum_{x=0}^{2^{n}-1}\sum_{y=0}^{2^{n}-1}\ket{I(x,y)}\ket{xy},
\end{equation}
where $\ket{I(x,y)}$ is a state of $q$ qubits which encodes the value of the pixel whose position $(x,y)$ is encoded in $\ket{xy}$ (an example is shown in Figure \ref{fig:NEQR-es2x2}).
\begin{SCfigure}[50][t] 
\centering
\begin{tikzpicture}[][]
	\fill[black] (-1,1) rectangle (0,0);
	\fill[black!50!white] (1,1) rectangle (0,0);
	\fill[black!50!white] (-1,-1) rectangle (0,0);
	\draw[step=1cm,black, line width=0.25mm] (-1,-1) grid (1,1);
	
	\node[text=white]     (0) at (-0.5, 0.5) {0};
	\node[]               (0) at (-0.5, -0.5) {128};
	\node[]               (0) at (0.5, 0.5) {128};
	\node[]               (0) at (0.5, -0.5) {255};
\end{tikzpicture}

\caption[]{Example of a $2\times2$ image whose encoding is $\ket{I}=\frac{1}{2}(\ket{0}\ket{00} + \ket{128}\ket{01} + \ket{128}\ket{10} + \ket{255}\ket{11})$
}
\label{fig:NEQR-es2x2}
\end{SCfigure} 

\subsubsection{Image retrieval}
By measuring the state \eqref{eq:NEQR-state} the result is $\ket{c_{xy}}\ket{xy}$. This result allows to state that the pixel of coordinates $(x,y)$ has an intensity value of $c_{xy}$. Since the pixel value is encoded in a further state $\ket{c_{xy}}$ entangled with the one encoding its position $\ket{xy}$, it is possible to recover exactly the pixel value with a single measurement. Other approaches, like FRQI and QPIE, which use the state amplitude to encode informations about the pixel, can only recover an approximation of it.

Therefore, the only difficulty lies in being able to make enough measurements to recover each states $\ket{I(x,y)}\ket{xy}$.
\\
\\
NEQR is better than FRQI when it is necessary to exactly recover the starting image after having performed some transformations on a quantum circuit. Moreover, as shown in \parencite{cit:zang}, the quantum state shown in \eqref{eq:NEQR-state} can be obtained with a quadratic speed-up as opposed to the state \eqref{eq:FRQI-state} used by FRQI.
\begin{figure}
\centering
\subfloat[][Originale]  {\includegraphics[width=.22\textwidth]{cat}} \quad 
\subfloat[][$N=\num{10 000}$]  {\includegraphics[width=.22\textwidth]{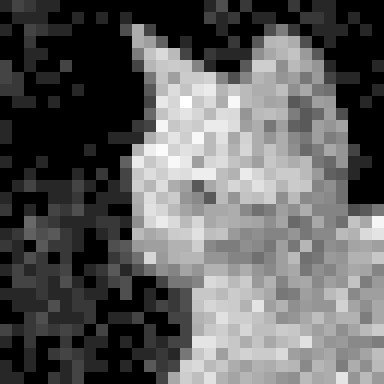}}\quad
\subfloat[][$N=\num{100 000}$] {\includegraphics[width=.22\textwidth]{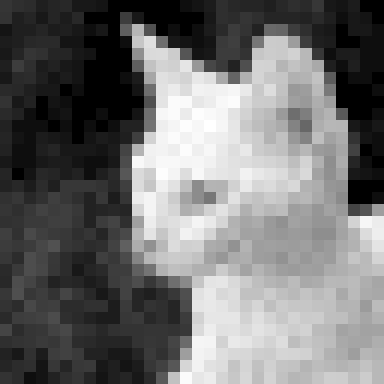}}\quad
\subfloat[][$N=\num{1 000 000}$]{\includegraphics[width=.22\textwidth]{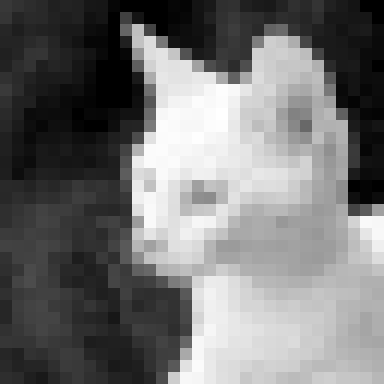}\label{fig:QIMR-lenna}}
\caption[]{Comparison between the original image \protect\footnotemark and images recovered with a different number of measurements from a quantum circuit}
\end{figure}

\footnotetext{Image taken from \href{https://pxhere.com/en/photo/999875}{pxhere.com} with licence \href{https://creativecommons.org/publicdomain/zero/1.0/deed.en}{CC0 1.0}}
\subsection{QPIE}
QPIE (\textit{Quantum Probability Image Encoding}) is an encoding proposed in \parencite{cit:yao}. An immediate advantage over the other encoding methods previously discussed is that it allows to encode rectangular $r\times c$ images. Moreover, the number of qubits necessary to encode the image are further reduced. This is particularly important given the scarcity of qubits available in today's quantum computers.

\subsubsection{Image encoding}
Given an $r\times c$ image I it can be turned into a vector by chaining its rows:
\begin{equation}\label{eq:QPIE-vec}
	I' = (I_{1,1}, I_{2,1},\dots,I_{r,1},I_{1,2},\dots,I_{r,c})^{T}.
\end{equation}
This vector, properly normalized, can be treated as the quantum state encoding the image in a quantum circuit; each state $\ket{i}$ represents the pixel position inside the image, while its amplitude encodes the intensity value of the pixel -- i.e. the value of the $i$-th element of $I'$.
\begin{equation}\label{eq:QPIE-stato}
	\ket{I} = \sum_{i=0}^{2^{n}-1}c_{i}\ket{i}\mbox{, } n = \lceil\log_{2}(rc)\rceil,
\end{equation}
where $c_{i} = I'(i)/||I'||$ is the normalized value of the $i$-th element of the vector shown in \eqref{eq:QPIE-vec}. It is important to note that, when working with rectangular images whose sides are not exact powers of 2, some of the states will not have a match with elements of $I'$; therefore a more precise formulation of $c_{i}$ is:
\begin{equation}
	c_{i} = \begin{cases}
				\frac{I'(i)}{||I'||} & \text{se } i<rc \\
				0                    & \text{otherwise}
			\end{cases}.
\end{equation}
This allows one to encode images of arbitrary size $r\times c$ using $n=\lceil\log_{2}(rc)\rceil$ qubits. Considering an image of size $2^{n}\times2^{n}$, as in previous examples, $2n$ qubits would be required.

\subsubsection{Image retrieval}\label{par:QPIE-recupero}
Despite having some advantages over the FRQI method, this encoding shares its problems regarding the extraction of the exact original image from its encoding quantum circuit. This is due to the fact that the pixel value is stored in the state amplitude, making it impossible to recover its exact value but only its approximation through many measurements. Given $N$ the total measurements taken and $N_{i}$ the times the $i$-th state was measured, the amplitude of $\ket{i}$ can be approximated as:
\begin{equation}\label{eq:QPIE-stima}
	P(\ket{i}) = c_{i}^{2} = \frac{N_{i}}{N}.
\end{equation}
As shown in Figure \ref{fig:QIMR-lenna}, $10^{6}$ measurements were necessary to recover a close approximation of a $32\times32$ image. With fewer measurements, the error in the approximation of the pixels' value is clearly visible.

\section{Quantum Edge Detection}\label{sec:qed}
QPIE is the quantum image representation method chosen to encode quantum images. While it is not optimal when an image must be retrieved from the quantum circuit, this method is particularly useful whenever this step is not required: for example, as shown later, it could be interesting to extract from the quantum circuit a percentage that indicates how similar two images are,
or to obtain a numerical value that tells if the image presents inversion symmetry. Moreover, a quantum edge detection algorithm based on QPIE was already proposed by \parencite{cit:yao}. After briefly discussing the idea behind this algorithm, an improved version is proposed.
\subsection{Quantum Edge Detection Algorithm}
\parencite{cit:yao} describes an algorithm to perform quantum edge detection with complexity $O(1)$. It performs edge detection by approximating the value of the first derivative of each pixel with the backward finite difference method. 

As described in \parencite{cit:yao}, by applying a single Hadamard gate to the first qubit of the quantum circuit encoding the image, the resulting state encodes the edge information in the amplitudes of the basis states of the form $\ket{\dots1}$:
\begin{align*}
	c_{0}\ket{0\dots00}  && \rightarrow && (c_{0}+c_{1})\ket{0\dots00}   \\
	c_{1}\ket{0\dots01}  && \rightarrow && (c_{0}-c_{1})\ket{0\dots01}   \\
	c_{2}\ket{0\dots10}  && \rightarrow && (c_{2}+c_{3})\ket{0\dots10}   \\
	c_{3}\ket{0\dots11}  && \rightarrow && (c_{2}-c_{3})\ket{0\dots11}   \\
	\dots 		     &&             && \dots					 
\end{align*}
As a result of the application of the Hadamard gate, all even states encode the differences $(c_{0}-c_{1})$, $(c_{2}-c_{3})$, $\dots$ -- i.e. the horizontal boundaries between pixels. This result can be easily shown on a $2^{n}\times2^{n}$ image encoded in a quantum state composed of $n$ qubits; the application of an Hadamard gate to the first qubit results in: 
\begin{gather}\label{equationlong}
	(\gateI^{\otimes n-1}\otimes \gateH)\ket{I} = \nonumber \\
	= (\gateI^{\otimes n-1}\otimes \gateH)(c_{0}\ket{0\dots00}+\dots+c_{2^{n}-1}\ket{1\dots11})  = \nonumber \\
	= c_{0}(\ket{0\dots0}\otimes\gateH\ket{0}) + \dots + c_{2^{n}-1}(\ket{1\dots1}\otimes\gateH\ket{1}) = \nonumber \\
	= c_{0}\ket{0\dots0}\otimes\frac{1}{\sqrt{2}}(\ket{0}+\ket{1})+ \dots + \nonumber \\
	   + \dots + c_{2^{n}-1}\ket{1\dots1}\otimes\frac{1}{\sqrt{2}}(\ket{0}-\ket{1}) 
\end{gather}
By factoring the common terms $\frac{1}{\sqrt{2}}(\ket{0}+\ket{1})$ and $\frac{1},{\sqrt{2}}(\ket{0}-\ket{1})$ it can be rewritten as:
\begin{gather}
	(c_{0}\ket{0\dots0} + \dots+c_{2^{n}-2}\ket{1\dots1})\otimes\frac{1}{\sqrt{2}}(\ket{0}+\ket{1}) + \nonumber \\
		+(c_{1}\ket{0\dots0} +\dots+c_{2^{n}-1}\ket{1\dots1})\otimes\frac{1}{\sqrt{2}}(\ket{0}-\ket{1}),
\end{gather}
lastly, by explicitly showing the result of the tensor products $\otimes\frac{1}{\sqrt{2}}(\ket{0}+\ket{1})$ and $\otimes\frac{1}{\sqrt{2}}(\ket{0}-\ket{1})$ the result is: 
\begin{gather}\label{eq:qed-stato}
	\frac{1}{\sqrt{2}}((c_{0}+c_{1})\ket{0\dots00} + (c_{0}-c_{1})\ket{0\dots01} + \dots +\nonumber\\
	+\dots+ (c_{2^{n}-2}-c_{2^{n}-1})\ket{1\dots11}).
\end{gather}

In order to obtain the remaining differences $(c_{1}-c_{2}), (c_{3}-c_{4}), \dots$ the original quantum state needs to be changed. A simple shift of the amplitudes, producing the state 
\begin{equation}\label{eq:QED-shift}
	\ket{I'} = c_{1}\ket{0\dots00} + \dots + c_{2^{n}-1}\ket{1\dots10} + c_{0}\ket{1\dots11}
\end{equation}
is enough to get the desired result after applying the Hadamard gate as described before.
This transformation can be represented by the following $2^{n}\times2^{n}$ matrix:
\begin{equation*}
	M_{\text{shift}} = 
	\begin{bmatrix}
	 0      & 0      & \dots  & 0      & 1      \\
	 1      & 0      & \dots  & 0      & 0      \\
	 0      & 1      & \dots  & 0      & 0      \\
	 \vdots & \vdots & \ddots & \vdots & \vdots \\
	 0      & 0      & \dots  & 1      & 0
	\end{bmatrix}.
\end{equation*}

\begin{figure}
\centering
	\begin{tikzpicture}
		\node[scale=0.7] {
		\begin{quantikz}
			\ket{0} & \gate[3]{Img_{1}} & \qw      & \qw       & \qw      & \qw      & \qw\dots & \swap{3} & \qw      & \qw      & \qw \\
			\dots   & \qw               & \swap{1} & \qw\dots        \\
			\ket{0} & \qw               & \targX{} & \gate{H}  & \qw      & \swap{3} & \qw\dots & \qw      & \qw      & \qw      & \qw \\
			\ket{0} & \gate[3]{Img_{2}} & \qw      & \qw       & \qw      & \qw      & \qw\dots & \swap{3} & \qw      & \qw      & \qw \\
			\dots   & \qw               & \swap{1} & \qw\dots         \\
			\ket{0} & \qw               & \targX{} & \gate{H}  & \qw      & \swap{1} & \qw\dots & \qw      & \qw      & \qw      & \qw \\
			\ket{0} & \qw               & \qw      & \qw       & \gate{H} & \ctrl{}  & \qw\dots & \ctrl{}  & \gate{H} & \meter{} & \qw  
		\end{quantikz}
		};
	\end{tikzpicture}
	\caption{Circuit used to compare two images after applying }
	\label{fig:case-study-circuit}
\end{figure}
\begin{figure}
\centering
\subfloat[][Borders from circuit (a)] {\includegraphics[width=.22\textwidth]{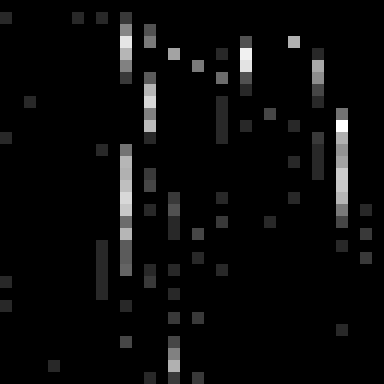}} \quad
\subfloat[][Borders from circuit (b)] {\includegraphics[width=.22\textwidth]{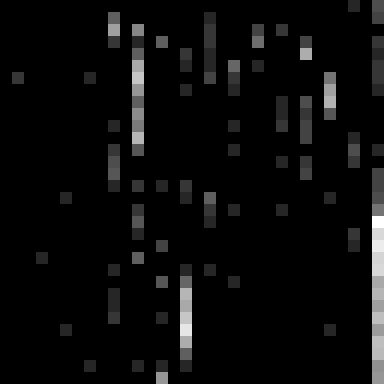}} \\
\subfloat[][Original image]         {\includegraphics[width=.22\textwidth]{cat}} \quad 
\subfloat[][Images (a) and (b) merged] {\includegraphics[width=.22\textwidth]{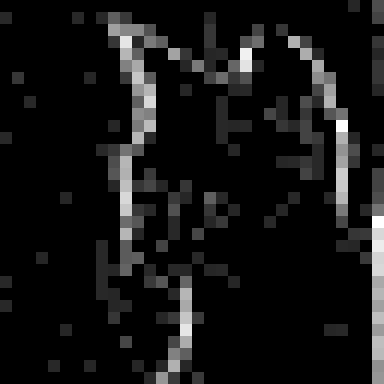}} \\

\caption[]{Result of the algorithm being applied to a $32\times32$ image, amplitudes were approximated by taking $\num{100 000}$ measurements of the same quantum circuit.}
\label{fig:QED-risultato}
\end{figure}
Figure \ref{fig:QED-circuiti} shows the quantum circuits implementing the two different circuits used to get edge information. Figure \ref{fig:QED-risultato} shows the results recovered from each circuit and an aggregated version.

\subsection{An improved Quantum Edge Detection Algorithm}
We present an improved version of the algorithm shown in the previous section. The idea is that by using the finite central difference --which produces better results in computing the first derivative of the pixels of an image-- a more accurate estimation of the boundary data can be obtained.

With a simple change in the original quantum state, it is possible to obtain the differences $(c_{0}-c_{2})$, $(c_{1}-c_{3})$, $\dots$ by doing so the first derivative (i.e. the boundaries informations) can be approximated more precisely. In order to get such result by applying an Hadamard gate to the first qubit of the quantum encoding state, the beginning state should be in the form:
\begin{align*}\label{eq:QED-better-state}
	\ket{I} & = c_{0}\ket{0\dots00} + c_{2}\ket{0\dots01} + c_{1}\ket{0\dots11} + \dots +\\ 
	 & +c_{2^{n}-2}\ket{1\dots10} + c_{2^{n}-1}\ket{1\dots11}.
\end{align*}

To obtain this state from an image encoded using the QPIE method it is necessary to apply a SWAP gate to its first two qubits, resulting in the desired swap of amplitudes of the states of the form $\ket{x01}$ and $\ket{x10}$. Then, by applying a single Hadamard gate to its first qubit the resulting state will encode the differences $(c_{0}-c_{2})$, $(c_{1}-c_{3})$, $\dots$ in its even amplitudes:
 \begin{align*}
	c_{0}\ket{0\dots00}  && \rightarrow && (c_{0}+c_{2})\ket{0\dots00} \\
	c_{2}\ket{0\dots01}  && \rightarrow && (c_{0}-c_{2})\ket{0\dots01} \\
	c_{1}\ket{0\dots10}  && \rightarrow && (c_{1}+c_{3})\ket{0\dots10} \\
	c_{3}\ket{0\dots11}  && \rightarrow && (c_{1}-c_{3})\ket{0\dots11} \\
	\dots            &&             && \dots                   
\end{align*}

\begin{figure}
\centering
	\subfloat[][Circuit to get the differences $(c_{0}-c_{2}), (c_{1}-c_{3}), \dots$] {
		\label{fig:detail}
		\begin{tikzpicture}
			\node[scale=0.70] {
			\begin{quantikz}
				\ket{0} & \gate[4]{Img} & \qw      & \qw       & \qw \\
				\dots   & \qw           & \qw      & \qw\dots  &      \\
				\ket{0} & \qw           & \swap{1} & \qw       & \qw \\
				\ket{0} & \qw           & \targX{} & \gate{H}  & \qw
			\end{quantikz}
			};
		\end{tikzpicture}
	} \quad 
	\subfloat[][Circuit to get the differences $(c_{2}~-~c_{4}), (c_{3}-c_{5}), \dots$] {
		\begin{tikzpicture}
			\node[scale=0.70] {
			\begin{quantikz}
				\ket{0} & \gate[4]{Img}  & \gate[4]{Shift_{2}} & \qw      & \qw      & \qw \\
				\dots   & \qw            & \qw                 & \qw      & \qw\dots &     \\
				\ket{0} & \qw            & \qw                 & \swap{1} & \qw      & \qw \\
				\ket{0} & \qw            & \qw                 & \targX{} & \gate{H} & \qw
			\end{quantikz}
			};
		\end{tikzpicture}
	}
	\caption{Circuits used to extract all edge information of the image}
	\label{fig:QED-circuiti-swap}
\end{figure}
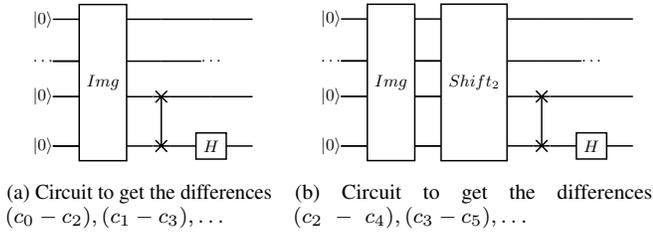

The quantum circuit implementing this algorithm is shown in Figure \ref{fig:detail}. The resulting complexity of the algorithm is still $O(1)$ --in terms of applied quantum gates-- therefore granting an exponential speed-up in comparison to any classical edge detection algorithm while providing a better approximation of the first derivative and boundaries information.

Similarly to the previous algorithm, in order to obtain the remaining differences between the amplitudes $(c_{2},c_{4}),(c_{3},c_{5}),$ $(c_{6},c_{8}),\dots$ it is necessary to apply a specific shift to the amplitudes of the starting state before applying the swap gate. In this case it is necessary to apply twice the shift transformation, obtaining the state:
\begin{equation}\label{eq:QED-better-state-shifted}
	\ket{I'} = c_{2}\ket{0\dots00} + c_{3}\ket{0\dots01} + \dots + c_{0}\ket{1\dots10} + c_{1}\ket{1\dots11}.
\end{equation}
This transformation can be described by the following $2^{n}\times2^{n}$ matrix:
\begin{equation*}
	M_{\text{shift}} = 
	\begin{bmatrix}
	 0      & 0      & \dots  & 1      & 0      \\
	 0      & 0      & \dots  & 0      & 1      \\
	 1      & 0      & \dots  & 0      & 0      \\
	 0      & 1      & \dots  & 0      & 0      \\
	 \vdots & \vdots & \ddots & \vdots & \vdots \\
	 0      & 0      & \dots  & 0      & 0
	\end{bmatrix}.
\end{equation*}

\begin{figure}
\centering
\subfloat[][Borders from circuit (a)] {\includegraphics[width=.22\textwidth]{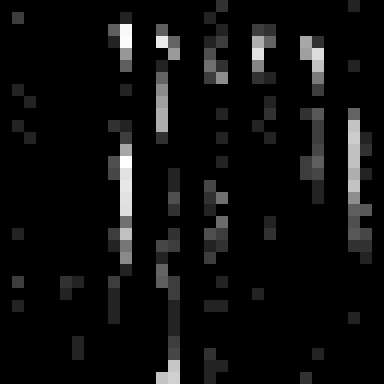}} \quad
\subfloat[][Borders from circuit (b)] {\includegraphics[width=.22\textwidth]{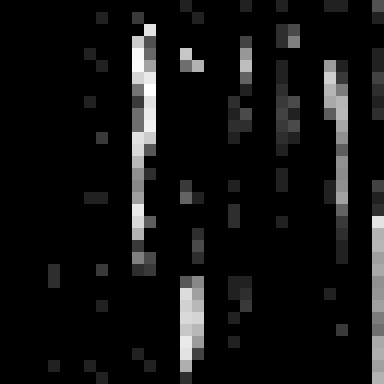}} \\
\subfloat[][Original image]         {\includegraphics[width=.22\textwidth]{cat}} \quad 
\subfloat[][Images (a) and (b) merged] {\includegraphics[width=.22\textwidth]{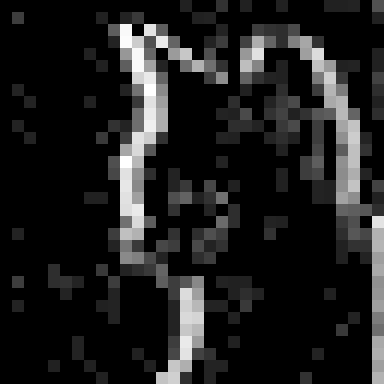}} \\

\caption[]{Result of the proposed algorithm being applied to a $32 \times 32$ image, amplitudes were approximated by taking $\num{100 000}$ measurements of the same quantum circuit.}
\label{fig:QED-better}
\end{figure}

As in the previous case it is necessary to use two distinct circuits to recover complete the complete edge information of the original image. The implementing circuits are shown in Figure \ref{fig:QED-circuiti-swap} while the result is shown in Figure \ref{fig:QED-better}

\section{A simple case study}\label{sec:case}
As stated previously QPIE is particularly suitable for those applications where it is not needed to recover the starting image. A simple case study to test the potential of this image encoding and quantum edge detection algorithm could be the comparison of images and their edges.

A swap test \parencite{cit:yao} can be performed to determine how similar the quantum states encoding the images are. As shown in Figure \ref{fig:case-study-circuit}, the circuit needs an additional qubit that will be measured yielding $0$ with a probability $P$ ranging from $1/2$ to $1$, indicating respectively two orthogonal or identical quantum states. The higher the similarity between the images encoded in the quantum circuit, the higher the probability that the measurement of the ancillary qubit will be $0$.
\\

\begin{figure}
\centering
\subfloat[] {\label{clothes-a}\includegraphics[width=.15\textwidth]{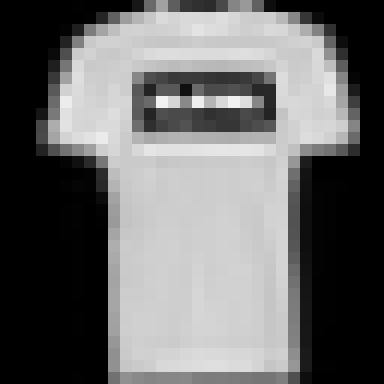}} \quad
\subfloat[] {\label{clothes-b}\includegraphics[width=.15\textwidth]{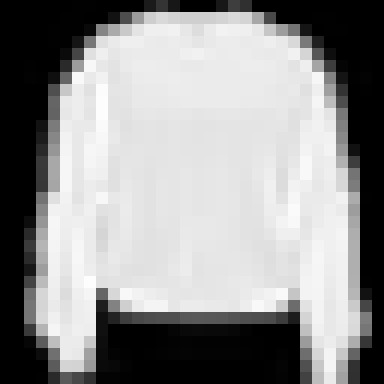}} \quad
\subfloat[] {\label{clothes-c}\includegraphics[width=.15\textwidth]{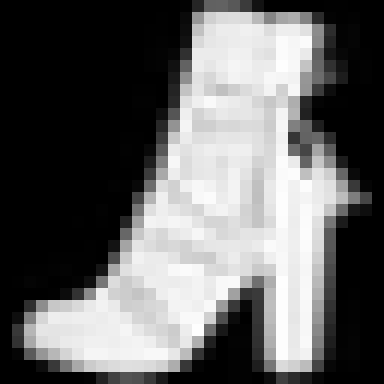}} \quad

\caption[]{The three compared images.}
\label{fig:case-study-imgs}
\end{figure}

The images used are taken from the Fashion-MNIST dataset \parencite{cit:xiao2017} and are shown in Figure \ref{fig:case-study-imgs}.

After comparing the images \ref{clothes-a} and \ref{clothes-b} the probability of measuring $0$ was estimated --by performing $1000$ measurements-- as $P_{ab}(0)=0.786$, when comparing their edges the estimated probability raised to $P_{ab\text{ edges}}(0)=0.812$. Image \ref{clothes-a} was also compared to image \ref{clothes-c}, obtaining the probabilities $P_{ac}(0) = 0.759$ and $P_{ac\text{ edges}}= 0.770$.
As expected, when the images are more similar, the probability $P$ of measuring $0$ is higher. Moreover, when comparing similar images --like \ref{clothes-a} and \ref{clothes-b}-- the similarity values are lower in comparison to the case where the edge of the images are compared: despite the fact that only half of the states --the ones of the form $\ket{\dots1}$-- encode the border information, the probability of measuring $0$ is higher. Indeed, the extracted edges are more informative of the peculiarity of each image allowing to obtain better comparisons. This can also explain why the comparison of the edges of \ref{clothes-a} and \ref{clothes-c} had a lower similarity score than the two images: the two images have notable differences accentuated by the edge extraction, thus resulting in an even lower similarity score.
\newpage

\section{Conclusions}\label{sec:concl}
To sum up, we considered a quantum edge detection algorithm --based on the QPIE image representation--, and proposed a better implementation using quantum properties to compute the first derivative using the finite mean difference approximation. This algorithm can perform edge detection in constant time --regardless of the size of the image-- providing an exponential speed-up compared to its classical counterparts, which would have to scan all the pixels of the image. 

Despite being a rather basic algorithm it shows how quantum computing could help the field of image processing in the future, despite the growing amount of information to be processed. 
However, there are many critical issues to address: first of all, freely accessible actual quantum computers --like IBM's-- still cannot offer enough qubits for proper experimentations on bigger images. Also, simulating quantum circuits quickly becomes unfeasible due to the overhead of the simulation which grows with the size of the circuit. 
\\
\\
As for future developments, an interesting step could be trying to implement a quantum version of a Sobel filter or a Prewitt filter for the QIMP representation. Such a filter would be far more useful than the simple one implemented in this article. However, as highlighted in \parencite{cit:yao}, finding a unitary matrix that would correspond to the application of such a filter is not an easy task.

\defbibheading{cartaceo}{\section*{References}}
\printbibliography[heading=cartaceo]

\end{document}